\documentclass{emulateapj}
\usepackage{apjfonts,graphics}

\newcommand{\simgt}{\lower.5ex\hbox{$\; \buildrel > \over \sim \;$}}
\newcommand{\simlt}{\lower.5ex\hbox{$\; \buildrel < \over \sim \;$}}

 \newcommand{\skaco}[1]{\langle{#1}\rangle}

\newcommand{\baredth}{\;\overline{\raise1.0pt\hbox{$'$}\hskip-6pt \partial}\;}
\newcommand{\edth}{\;\raise1.0pt\hbox{$'$}\hskip-6pt\partial\;}

\lefthead{Broadhurst, Takada, Umetsu et al.}
\righthead{Steep Mass Profile of A1689}
\begin{document}

\title{The Surprisingly Steep Mass Profile of Abell 1689, from a 
Lensing Analysis of Subaru Images$^1$
}
\author{Tom Broadhurst\altaffilmark{2}, 
Masahiro Takada\altaffilmark{3}, 
Keiichi Umetsu\altaffilmark{4}, 
Xu Kong\altaffilmark{5}, 
Nobuo Arimoto\altaffilmark{5}, 
Masashi Chiba\altaffilmark{3},
Toshifumi Futamase\altaffilmark{3}}
\altaffiltext{1}{Based in part on data collected at the Subaru Telescope,
  which is operated by the National Astronomical Society of Japan}
\altaffiltext{2}{School of Physics and Astronomy, Tel Aviv University, Israel}
\altaffiltext{3}{Astronomical Institute, Tohoku University, Sendai 980-8578, Japan}
\altaffiltext{4}{Institute of Astronomy and Astrophysics, Academia
Sinica,  P.~O. Box 23-141, Taipei 106,  Taiwan, Republic of China}
\altaffiltext{5}{National Astronomical Observatory of Japan, Mitaka 181-8588, Japan}

\begin{abstract}
Subaru observations of A1689 $(z=0.183)$ are used to derive an accurate,
model-independent mass profile for the entire cluster, $r\simlt
2{\rm Mpc}/h$, by combining magnification bias and distortion
measurements.  The projected mass profile steepens 
quickly
with increasing radius, falling away to zero at $r\sim 1.0{\rm
Mpc}/h$, well short of the anticipated virial radius.  Our profile
accurately matches onto the inner profile, $r\simlt 200{\rm kpc}/h$,
derived from deep HST/ACS images. The combined ACS and Subaru
information is well fitted by an NFW
profile with virial mass, $(1.93\pm 0.20)\times 10^{15}M_\odot$, and 
surprisingly high concentration, $c_{\rm
vir}=13.7^{+1.4}_{-1.1}$, significantly larger than theoretically expected
($c_{\rm vir}\simeq 4$), corresponding to a relatively steep overall
profile.  A slightly better fit is achieved with a steep power-law
model, 
$d\log \Sigma(\theta)/d\log \theta\simeq -3$, 
with a core $\theta_c\simeq 1.\!' 7$
($r_c\simeq 210{\rm
kpc}/h$), 
whereas an isothermal profile is strongly rejected.  
These results are based on a reliable sample of background galaxies
selected to be redder than the cluster E/S0 sequence. By including the
faint blue galaxy population a much smaller distortion signal is
found, demonstrating that 
blue cluster members significantly dilute the true signal for $r\simlt
400{\rm kpc}/h$. This contamination is likely to affect most weak
lensing results to date.
\end{abstract}                   


\section{Introduction}\label{section1}

Numerical simulations based on the cold dark matter (CDM) scenario are
reliable enough to make statistical predictions for the mass profiles
of clusters.  
The gradient of an `NFW' profile is predicted to
monotonically steepen with increasing radius (Navarro, Frenk \&
White 1997),  with logarithmic slopes shallower than an isothermal
profile interior to the characteristic radius $r<r_s$, but steeper at
larger radius, approaching to $r^{-3}$ at $r\rightarrow r_{\rm vir}$.
This curvature is particularly pronounced for massive clusters, where
the halo is expected to have a relatively low concentration, providing a
clear prediction. Weak lensing work has yet to make a definitive
statement regarding the validity of the NFW profile, with the 
analysis claiming only broad consistency with both the singular
isothermal case and the NFW model (Clowe \& Schneider 2001; Bardeau et
al. 2004). 
Recently, Kneib et al. (2003) made a more comprehensive lensing study
using the panoramic sparse-sampled HST/WFPC2 data of Cl0024, showing that at large
radius, the mass profile is steeper than the isothermal 
case, consistent with an NFW profile for $r>r_s$.
A firmer
constraint is possible with magnification information, breaking the
mass-sheet degeneracy inherent to distortion measurements (Broadhurst,
Taylor \& Peacock 1995, hereafter BTP). Detections of a number-count
depletion have been claimed for several clusters 
(Fort et al. 1997; Taylor et al. 1998; 
Athreya et al.  2002;
Dye et al. 2002).

Recently, the central region of A1689 ($z=0.183$) has been imaged in
detail with the Advanced Camera for Surveys (ACS), revealing 106 multiply lensed
images of 30 background galaxies (Broadhurst et al. 2004, B04). Many
radially directed images define a radial critical curve, inside which
small counter-images are identified, so the mass profile can be traced
in detail to the center of mass.  An NFW profile fits well over the
restricted ACS field, $r\simlt 200 {\rm kpc}/h$, but with a somewhat
larger concentration, $c_{\rm vir}=8.2^{+2.1}_{-1.8}$, than expected for
massive clusters, $c_{\rm vir}\sim 4$ (e.g., Bullock et al. 2001). To
examine the shape of the full profile, we turn to the wide-field prime
focus camera, Suprime-Cam, of the $8.2$m Subaru telescope. Suprime-Cam
provides an unparalleled combination of area ($34'\times 27'$) and
depth (Miyazaki et al. 2002). We measure the lensing distortion and
magnification of background red galaxies and combine them to derive a
model-independent mass profile out to $r\sim 2{\rm Mpc}$, allowing a
definitive comparison with model profiles. Throughout this {\it
Letter}, the concordance $\Lambda$CDM cosmology is adopted
($\Omega_{\rm m0}=0.3$, $\Omega_{\lambda 0}=0.7$, $h=0.7$). 
Note that one arcminute corresponds to the physical scale
$129$kpc$/h$  for this cluster.

\section{Data reduction and Sample selection} 
\label{data}

Suprime-Cam imaging data of A1689 in $V$ (1,920s) and SDSS $i'$ (2,640s) 
were
retrieved from the Subaru archive,
SMOKA.
Reduction software developed by Yagi et al (2002)
 was used for flat-fielding,
instrumental distortion correction, differential refraction, PSF
matching, sky subtraction and stacking.  The resulting FWHM is
$0\arcsec\!\!.82$ in $V$ and 
$0\arcsec\!\!.88$ in $i'$ with $0\arcsec\!\!.202$
pix$^{-1}$, covering a field of $30'\times 25'$.

Photometry is based on a combined $V+i'$ image using 
SExtractor (Bertin \& Arnaut 1996).  The limiting magnitudes are
$V=26.5$ and $i'=25.9$ for a $3\sigma$ detection within a $2$\arcsec
aperture (we use the AB magnitude system). 
For the number counts to measure magnification, we define a sample of 
8,907 galaxies ($12.0$ arcmin$^{-2}$) with $V-i'>1.0$. 
For distortion measurement, 
we define a sample of 5,729 galaxies ($7.59$ arcmin$^{-2}$)
with colors $0.22$ mag redder than the color-magnitude sequence of
cluster E/S0 galaxies, $(V-i') +0.0209i'-1.255 > 0.22$. The smaller
sample is due to the fact that distortion analysis requires galaxies
used are well resolved to make reliable shape measurement. 
We adopt a limit of $i'<25.5$ 
to avoid incompleteness effect. 
Our red galaxies are very reasonably
expected to lie behind the cluster, made redder by larger
k-corrections. The counts of all galaxies including the dominant faint
blue population is much larger, $\sim 40$ arcmin$^{-2}$. However, as we
show below, the distortion signal of the full sample is significantly
diluted
 within
$r\simlt 400{\rm kpc}/h$, compared to the red sample, indicating that blue
cluster members
contaminate the sample. The mean redshift of the red
galaxies 
is estimated to be
$z_s\simeq1\pm0.1$, based on deep photo-z estimation for deep field
data (Benitez et al. 2002). In the following we will assume
$\skaco{z_s}=1$ for the mean
redshift, but note that the low redshift of A1689 means that for
lensing work, a precise knowledge of this redshift is not critical.

\section{Lensing Distortions}
\label{shear}

We use the IMCAT package developed by N. Kaiser for our distortion
analysis, following the well tested formalism outlined in Kaiser,
Squires \& Broadhurst (1995), with modifications described by Erben et
al.(2001).

Briefly, to estimate the lensing distortion from the observed
ellipticities of galaxy images, we first correct for PSF anisotropy using
the available stars.  The mean stellar ellipticity before correction is
$(\bar{e_1}^*, \bar{e_2}^*) \simeq (-0.013, -0.018)$
 over the survey field, and is reduced by the correction to only
$ {\bar{e}^{*{\rm res}}_1} = (0.47\pm 1.32)\times 10^{-4}$ and $
{\bar{e}^{*{\rm res}}_2} = (0.54\pm 0.94)\times 10^{-4}$.  We also
correct for the isotropic smearing effect caused by seeing as well as by
 the
window function used in the shape estimate. Full details will be
presented in Umetsu et al. (2004, in preparation) including the 2D
maps.

%%%%%%%%%%%%%%%%%%%%%%%%%%%%%%%%%%%%%%%%%%%%%%%%%%%%%%%%%%%%%%%%%%%%%%%%%
\begin{figure}
\epsscale{1.15}
\plotone{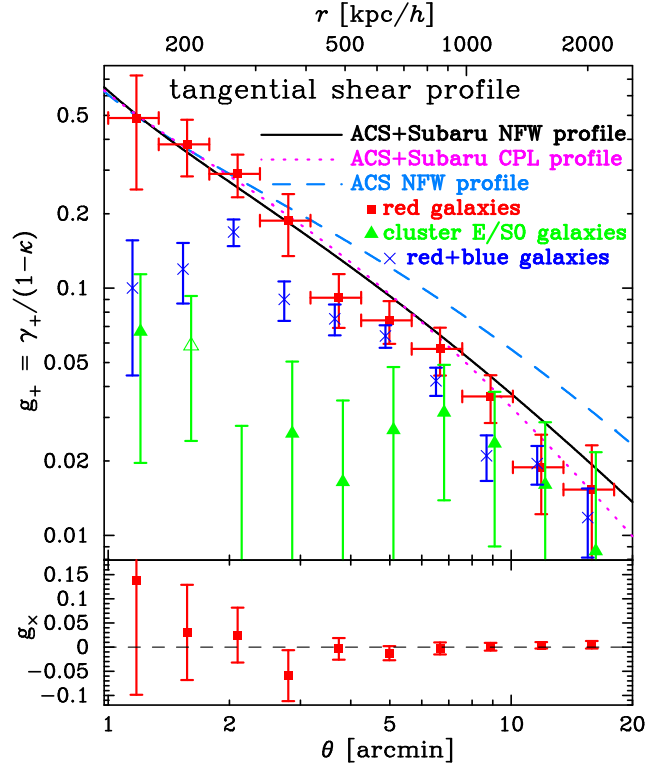}
\caption{Radial profiles of the tangential shear (upper panel) and the
$45^\circ$ rotated ($\times$) component (lower panel). The square symbols
correspond to the red background sample.  The cross symbols correspond
to the full sample of faint red+blue galaxies, excluding objects falling
on the narrow cluster E/S0 sequence. The distortion signal of the full
sample falls well below that of the background red galaxies for
$r\simlt 400{\rm kpc}/h$, indicating significant contamination by blue
cluster members.
The triangle symbols denote galaxies with photometry coincident with the 
color-magnitude sequence of cluster E/S0 galaxies (the open 
 symbol indicates the negative signal).
The solid and dotted curves show 
an NFW profile with a large
concentration, $c_{\rm vir}\approx 14$, and a power-law profile with 
slope $n=3.16$ and core $\theta_c=1.65 $ arcmin (see \S\ref{mass}), 
 respectively, both matching well the overall profile of the red galaxy sample. 
The best-fitting model (dashed curve) 
based on the inner ACS measurements (B04), with
$c_{\rm vir}\simeq 8$, clearly overestimates the outer profile.}
\label{fig:shear}
\end{figure}
%%%%%%%%%%%%%%%%%%%%%%%%%%%%%%%%%%%%%%%%%%%%%%%%%%%%%%%%%%%%%%%%%%%%%%
Figure \ref{fig:shear} shows the radial profiles of the tangential
distortion, $g_{+}$, and the 45-degree rotated ($\times$) component,
$g_{\times}$,
 for
the different samples of galaxies, where the cluster center is well
determined by the locations of brightest cluster galaxies and/or giant arcs. 
Note that lensing induces only
the tangential shear in the weak lensing regime.
The error bars  are 68\% confidence intervals computed from the bootstrap
resampled data.
It is clear that
the observed signal is significant to the limit of our
data, $\sim 20'$ or $2$Mpc$/h$, 
and that the $\times$-component of the
red galaxy sample is consistent with a null signal at all radii,
indicating the reliability of our distortion analysis.

One can see that the NFW prediction for source redshift $z_s=1$ 
($c_{\rm vir}=8.2$ and $M_{\rm vir}=2.6\times 10^{15}h M_\odot$), which
fits best the ACS strong lensing data restricted to
the central region $\simlt 2'$
(B04), matches onto the Subaru distortions at $\simlt 3'$ for the background red
galaxies, but increasingly overestimates the distortions at larger
radii $\simgt 3'$. A steeper NFW profile with $c_{\rm vir}\sim 14$ 
(solid curve) better reproduces the distortion profile at all radii, as
does a cored power-law profile (dotted curve),
(see \S\ref{mass} for the details).

A careful background selection is critical. If we select all galaxies
irrespective of color with magnitude limit $i'<25.5$, and exclude
only the cluster sequence galaxies, then we find that the distortion
signal falls below the red background sample by a factor of 2-5 at
$r\simlt 400{\rm kpc}/h$. 
While the slight dilution within a factor of 2
is still apparent at large radii $\simgt 5'$, 
blue
cluster members
must be significantly contaminating the full sample at the small
radii, reducing the distortion signal. 
 This full sample selection is very similar to Bardeau et al. (2004)
and Clowe \& Schneider (2001), and we find close agreement. Hence it
is apparent that these analyses underestimate the true distortion
signal and explains why they underpredicted the Einstein radius, $\sim
20$\arcsec, compared with the much larger observed radius of $\sim
45''$ (for $z_s=1$), 
a puzzle noted by Clowe \& Schneider (2001).

It is also instructive to examine the tangential distortions of
galaxies whose photometry places them in the limits bounding the cluster
sequence. %The signal is consistent with a null signal at $\simlt 5'$
A null
signal is found for $\simlt 5'$ 
as expected, where the cluster member is
prominent above the background. But a positive distortion, with weaker
signal than the red galaxy sample, is found
for $ \simgt 5'$,
indicating that background, lensed galaxies contribute significantly.

\section{Magnification Bias}\label{mag}
%%%%%%%%%%%%%%%%%%%%%%%%%%%%%%%%%%%%%%%%%%%%%%%%%%%%%%%%%%%%%%%%%%%%%%%%%
\begin{figure}
\epsscale{1.15}
\plotone{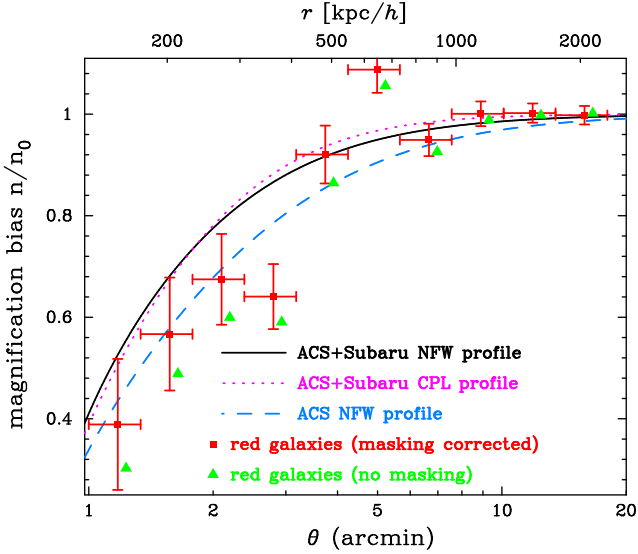}
\caption{Number-count profile of the red background galaxies (square
 symbols). The counts are normalized by the unlensed counts estimated
 from the cluster outer region (see text for details). 
The triangle symbols show the counts without the mask
 correction due to cluster members. The model curves are shown for
 comparison as in Figure \ref{fig:shear}.} 
\label{fig:magbias}
\end{figure}
%%%%%%%%%%%%%%%%%%%%%%%%%%%%%%%%%%%%%%%%%%%%%%%%%%%%%%%%%%%%%%%%%%%%%%

Lensing magnification, $\mu(\vec\theta)$, influences the observed
surface density of background galaxies, expanding the area of
sky, and enhancing the flux of galaxies 
(BTP). 
The number counts for a given magnitude cutoff $m_{\rm cut}$, 
approximated as a power law with slope $s\equiv
d\log N_0(m)/dm$, are modified as
$N(<m_{\rm cut})=N_0(<m_{\rm cut})\mu^{2.5s-1}$, where $N_0$ is the
unlensed counts.  
Thanks to the large Subaru FOV, 
the normalization and slope of $N_0$ for
our red galaxy sample are reliably estimated as $n_0=12.6 \pm 0.23$ arcmin$^{-2}$ and
 $s=0.22\pm0.03 $ from the outer region 
$\ge 10'$. The slope is less than the lensing invariant slope, $s=0.4$, 
so a net deficit of background galaxies is expected.

We conservatively account for the masking of observed sky
by excluding a generous area $\pi a b$ around each cluster
sequence galaxy, where $a$ and $b$ are defined as 3 times
the major ({\tt A\_IMAGE}) and minor axes ({\tt B\_IMAGE}) computed
from SExtractor, corresponding
to the isophotal detection limit (Bertin \& Arnouts 1996).  
The number density in each radial annulus
is then calculated by excluding this
area and simply renormalising. 
Note that, if we adopt the masking
factor of $2$ or $4$ instead of $3$, the results shown below are little
changed. 

Figure \ref{fig:magbias} shows that
the red galaxy counts 
are clearly depleted, % as expected, given their shallow slope,
with a clear trend towards higher
magnification in the center. The
masking area is negligible at large radius and rises to $\sim 20\%$
of the sky close to the center,  $r \simlt 3'$. 
Comparison with models shows 
that the
data are broadly in agreement with the profiles as described 
in \S\ref{mass}. We have ignored the intrinsic clustering of
background galaxies, which seems a good approximation, though some
variance is apparent in 2D maps (Umetsu et al. 2004) and may explain
a discrepant point at $\sim 5'$ in the radial profile.

\section{Model-Independent Mass Profile}\label{mass}

The relation between distortion and convergence is non-local, and
masses derived from distortion data alone suffers from a mass sheet
degeneracy. However, by combining the distortion and magnification
measurements the convergence can be obtained unambiguously. Here we
derive a model-independent, discrete convergence profile in 10
logarithmically spaced bins for $1'\le\theta\le
18'$: $\kappa_i\equiv \kappa(\theta_i)$ for $i=1,2,\dots, 10$,
representing 10 free parameters which are constrained with 20 data
points in Figures \ref{fig:shear} and \ref{fig:magbias}.

To perform the mass reconstruction, we need to express the lensing
observables in terms of the binned convergence profile.
The tangential shear amplitude in the $i$-th radial bin
can be expressed as
$\gamma_{+i}=\bar{\kappa}_i-\kappa_i$ (Fahlman et al.
1994), where $\bar{\kappa}_i$ is the average convergence
interior to radius $\theta_i$ and expressed as
$\bar{\kappa}_i\equiv \bar{\kappa}(\theta_i)=2/\theta_i^2\sum_{j=1}^i 
\theta_j^2 \kappa_j h$ with the bin width $h\equiv \Delta
\ln\theta$.
Once the shear profile is given, it is straightforward to compute
the binned tangential distortion and 
magnification via the relations 
$g_{+i}=\gamma_{+i}/(1-\kappa_i)$ (at radii of our interest) and 
$\mu_i=|(1-\kappa_i)^2-\gamma_{+i}^2|^{-1}$.  % (Broadhurst 1995).  
However, for the discrete model, to properly 
compute the shear in the first bin,
we need to specify the 
mass interior to radius $1'$, which is readily obtained from 
the well constrained ACS
derived profile. We have checked that, if we instead adopt the 
model independent mass interior to the
Einstein radius ($\approx 45$\arcsec ${}$),
the results are very similar.

The best-fitting model $\kappa_i$ is then derived by properly weighting
signal-to-noise ($S/N$) ratios of the distortion and magnification data
(e.g., see Schneider, King \& Erben 2000).
%, where
%the distortion $S/N$ is higher than the magnification.
 Since covariance between 
the distortion and magnification signals can be ignored,
the $\chi^2$ is simply expressed as
$\chi^2=\chi^2_g+\chi^2_\mu$ with
$\chi_g^2\equiv \sum_{i=1}^{10}
\left[g_{+i}^{\rm model}-g_{+i}^{\rm obs}\right]^2/\sigma_{gi}^2$ and
$\chi_\mu^2\equiv \sum_{i=1}^{10}
\left[N_i^{\rm model}-N_{i}^{\rm obs}\right]^2/N^{\rm obs}_i.$
Note that $N_i$ denotes the number counts in the $i$-th radial annulus and the
model counts $N_i^{\rm model}$
is given in terms of the unlensed surface density $n_0$ and
the model magnification as $N_i^{\rm model}=\mu_i^{2.5s-1}n_{0}\Omega_i$,
where $\Omega_i$ is the effective observed area of the $i$-th annulus excluding the
masking area by the cluster members (see \S\ref{mag}).  

%%%%%%%%%%%%%%%%%%%%%%%%%%%%%%%%%%%%%%%%%%%%%%%%%%%%%%%%%%%%%%%%%%%%%%%%%
\begin{figure*}[t]
\epsscale{1.2}
%\epsscale{.95}
\plotone{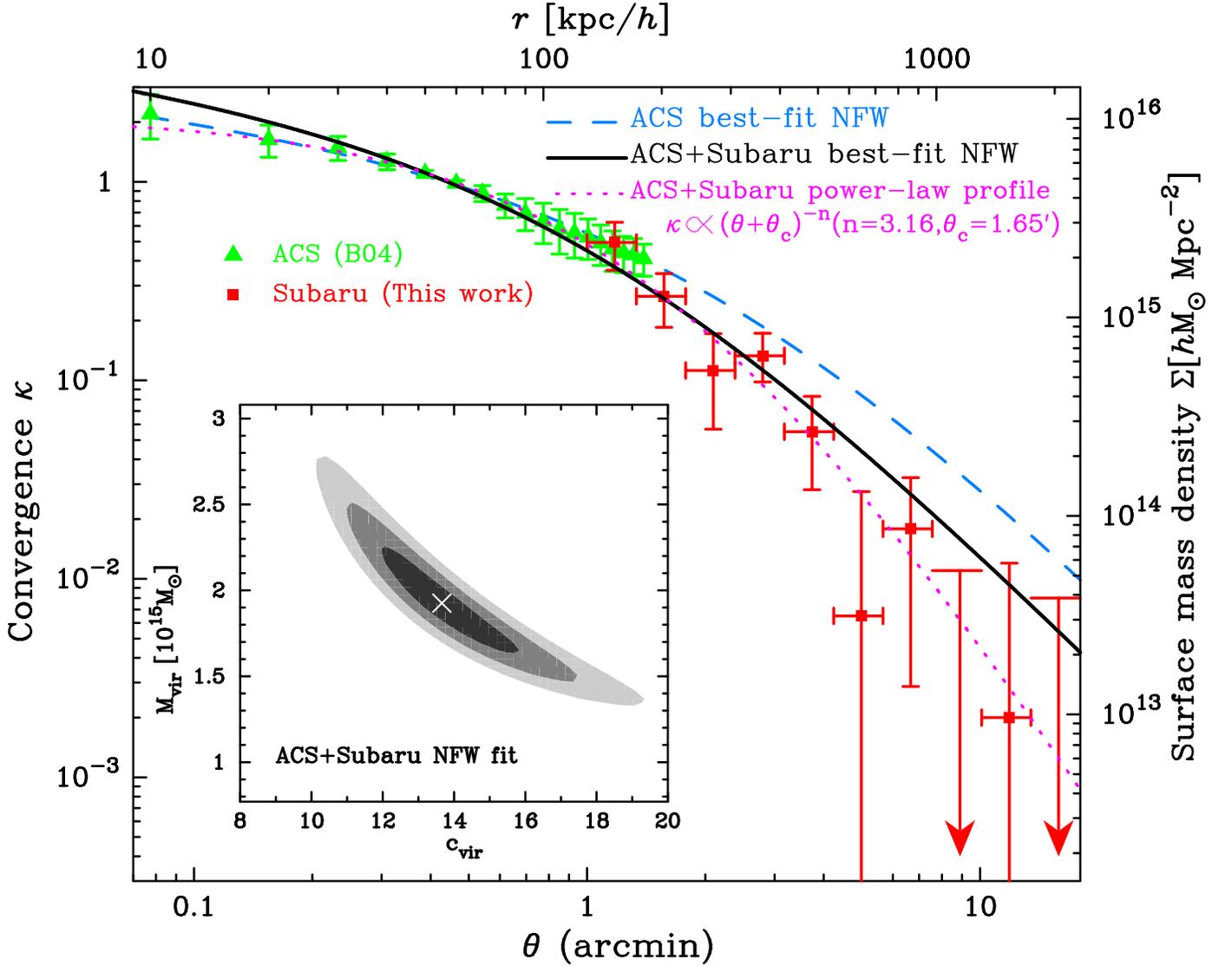}
\caption{Reconstructed mass profile. The triangle and square symbols
 with error bars show the results from the ACS strong lensing
 analysis (B04) and the Subaru weak lensing analysis (this work), 
respectively. The
 dashed and solid curves show the best-fitting NFW profiles for the ACS
 data alone and for the combined ACS+Subaru profile, respectively. The
 best-fitting NFW profile for the ACS+Subaru profile
 has a high concentration, $c_{\rm vir}=13.7$,
 and somewhat overestimates the inner slope and is a bit
 shallow at large radius. %, but has a very similar form to the data,
 %increasingly steepening with radius. 
A cored power-law provides a
 better
 fit to the full projected mass profile. The inset plot
 shows the $68\%$, $95\%$ and 99.7\% confidence 
levels in the ($c_{\rm vir}$, $M_{\rm vir}$) plane for the ACS+Subaru
 NFW fitting ($\Delta\chi^2=2.3$, $6.17$ and $11.8$).}
\label{fig:conv}
\end{figure*}
%%%%%%%%%%%%%%%%%%%%%%%%%%%%%%%%%%%%%%%%%%%%%%%%%%%%%%%%%%%%%%%%%%%%%%
The square symbols in Figure \ref{fig:conv} show the best-fitting mass
profile obtained from the Subaru data, where the minimized $\chi^2$
value is $\chi_{\rm min}^2/{\rm dof.}=27.9/10$ for 10 degrees of
freedom. The bad $\chi^2_{\rm min}$ is mainly due to the two data
points at $\sim 3'$ and $5'$ in Figure \ref{fig:magbias}, which
apparently deviate from a continuous curve and are probably affected
by the intrinsic clustering of background galaxies. Excluding the two
data points leads to an acceptable fit $\chi^{2}_{\rm min}/{\rm
dof.}=12.4/8$, meaning the best-fit mass profile well reproduces the
distortion and magnification data 
simultaneously\footnote{If we use the distortion data for the red+blue galaxies (the
cross symbols in Figure \ref{fig:shear}), the minimum $\chi^2$ gets
worse, $\chi^2_{\rm min}/{\rm dof.}=147/10$, and the mass profile is
significantly underestimated.}. 
The outer profile at $\theta \simgt 9'$ 
 is consistent with a null signal to
within $1\sigma$ (the points at $\sim 9'$ and $16'$ are $1\sigma$
upper limits), implying a more rapid decay in mass profile than
expected from the standard NFW profile, falling well short of the
anticipated virial radius. We have also checked that this result is
not sensitive to varying the count-slope and the unlensed background
galaxy density within the $1\sigma$ measurement errors.

Force fitting an NFW profile 
to the Subaru mass profile yields
$M_{\rm vir}=1.69^{+0.30}_{-0.28}\times 10^{15}M_{\odot}$  ($r_{\rm
vir}=1.95\pm 0.11 $Mpc/$h$) for the virial mass (radius), 
where $\chi^2_{\rm min}/{\rm dof.}=5.36/8$.
Note that the NFW profile is specified by the two parameters,
$M_{\rm vir}$ and halo concentration $c_{\rm vir}$ (Bartelmann 1996),
and we have adopted the flat prior
$c_{\rm vir}\le 30$ because the NFW profiles with $c_{\rm vir}\simgt
20$ can not be distinguished by the Subaru data alone due to lack of
information on the inner density profile.  The
error quotes 68$\%$ confidence intervals ($\Delta\chi^2\equiv
\chi^2-\chi^2_{\rm min}\le 1$).
A pure power-law
profile given by $\kappa\propto \theta^{-n}$ also provides a
good fit with $n=2.08^{+0.30}_{-0.27}$
($\chi^2_{\rm min}=4.07/8$), corresponding to a 3D profile, $\rho
\propto r^{-3.08}$. 
A singular isothermal sphere with $n=1$ is excluded at more than $4\sigma$ level

Next, we consider the combined Subaru and ACS profile, where the ACS profile
with error bars is taken from Figure 22 in B04 and the amplitude is
scaled to 
$z_s=1$ from the $z_s=3$ result. 
By combining the strong and weak lensing analyses, we can trace the
mass distribution over a large range in amplitude
$\kappa\sim [10^{-3},1]$ and in radius $r=[10^{-2},2]$Mpc$/h$.  
In this case, we
have 22 independent data points in total excluding the ACS data
points at radii overlapping with the Subaru data.  
The best-fitting NFW profile is given by
$c_{\rm vir}=13.7^{+1.4}_{-1.1}
$ and $M_{\rm
vir}=(1.93\pm 0.20)\times 10^{15}M_\odot$ ($r_{\rm vir}=2.04
\pm0.07$Mpc/$h$), 
and has an acceptable fit $\chi^{2}_{\rm
min}/{\rm dof.}=13.3/20$. 
The inset plot shows how the constraints are
degenerate in the ($c_{\rm vir}$, $M_{\rm vir}$) plane. 
Hence, this result gives the lensing
based confirmation of the NFW profile over the radii
we have considered, however, 
the mass
distribution appears to be much more concentrated toward the center
than the CDM simulations predict for a halo of the above mass,
$c_{\rm vir}=4.0$ (Bullock et al. 2001). A generalized NFW profile given by
$\rho\propto r^{-1.5}(1+r/r_s)^{-1.5}$ (e.g., Moore et al. 1998)
is disfavored  ($\chi^2_{\rm min}/{\rm dof.}=28.3/20$; see Umetsu et
al. 2004 for more details), being too steep in the center. 
A cored power-law profile, $\kappa\propto (\theta+\theta_c)^{-n}$,
gives a better fit:
$\chi_{\rm min}^2=4.49/19$ with
 a steep slope, 
$n=3.16^{+0.81}_{-0.72}$,
and a core of $\theta_c=1.65^{+0.77}_{-0.61}$ arcmin ($r_c=214^{+99}_{-78}$
kpc/$h$). 
A softened isothermal profile is
strongly rejected ($10\sigma$ level!). 

\section{Discussion and Conclusions}
We have obtained a secure, model-independent mass profile of A1689
over $1'<\theta\simlt 20'$ ($100{\rm kpc}/h \simlt r\simlt {\rm
2Mpc}/h$) by combining the distortion and magnification-bias
measurements from high-quality Subaru imaging. We have seen that to
reliably measure distortions it is critically important to securely
select background galaxies in order to avoid dilution of the
distortion signal by blue cluster members and foregrounds (see Figure
\ref{fig:shear}). Thus we have resolved the discrepancy between the
small Einstein radius ($\sim 20$\arcsec) inferred from the previous
work based on largely monochromatic measurements and the observed
radius ($\sim 45$\arcsec). The mass profile of A1689 obtained from the
Subaru and ACS data covers 2 orders of magnitude in radius,
$[10^{-2},2]{\rm Mpc}/h$, and shows a continuously steepening profile
with increasing radius, very similar to an NFW profile but with a much
higher concentration than expected.  The best-fitting NFW profile has
$c_{\rm vir}\simeq 14$, significantly larger than expected $c_{\rm
vir}\simeq 4$, corresponding to the profile expected for a much lower
mass halo of $\sim 10^{11-12}M_{\odot}$ (Bullock et al. 2001), 3-4
orders of magnitude less than the mass of A1689.  A higher
concentration may imply a higher than expected redshift for cluster
formation, corresponding to a greater mean cosmological mass density,
so that collapsed objects of a given mass have a higher internal
density. A higher redshift of cluster formation may also help account
for the lack of evolution observed in the properties of galaxy
clusters (Mathis, Diego \& Silk 2004).
Although A1689 is a very round shaped cluster with evidence of only
modest substructure (B04; also see Andersson \& Madejski 2004 for 
related discussion based on the $X$-ray data), 
projection of structure along the line of
sight may potentially boost the surface density increasing the derived
concentration. Substructure may be examined further with redshift
measurements for this cluster. Simulations may also provide a good
guide to the level of bias one may expect due to projection in general
(e.g., Clowe, Lucia \& King 2004).

Careful lensing work on other
clusters may also point to higher concentrations than expected.
 Gavazzi
et al. (2003) quote $c\sim 12$ for the cluster MS2137-23, but with a
low significance. The situation for Cl0024 might be complicated by
prominent substructure, with a high concentration for the main
clump, $c\simeq22$, and a low concentration, $c\simeq 4$, for the secondary
clump (Kneib et al. 2003).
Subaru imaging in more passbands will provide many reliable
photometric redshifts for refining the background selection, helping
to improve the distortion signal and allowing the lens magnification
to be separated from the background clustering. In addition, the
redshift dependence of the distortion and magnification signals may
be extracted to constrain cosmological parameters.  

\acknowledgments We thank R.~Barkana, D.~Clowe 
B.~Jain, D.~Maoz, E.~Ofek, Y.~Suto, and I.~Tanaka for
valuable discussions,  and especially thank anonymous referee
for useful comments which led to improvement on the
manuscript.
We are also grateful to N.~Kaiser
for making the IMCAT package publicly available.  This work is
partially supported by the COE program at Tohoku
University.  XK thanks the JSPS for support and 
TJB thanks the generosity of the JSPS and the
hospitality of NAOJ and the Astronomical Institute of Tohoku
University.

\end{document}